\documentclass[11pt]{article}
\usepackage{amsmath}
\usepackage{amsfonts,amssymb,latexsym}

\begin{document}

\newcommand{\nl}{\nonumber\\}
\newcommand{\nnl}{\nl[6mm]}
\newcommand{\nle}{\nl[-2.5mm]\\[-2.5mm]}
\newcommand{\nlb}[1]{\nl[-2.0mm]\label{#1}\\[-2.0mm]}

\renewcommand{\leq}{\leqslant}
\renewcommand{\geq}{\geqslant}

\renewcommand{\theequation}{\thesection.\arabic{equation}}
\let\ssection=\section
\renewcommand{\section}{\setcounter{equation}{0}\ssection}

\newcommand{\be}{\bes}
\newcommand{\ee}{\ees}
\newcommand{\bes}{\begin{eqnarray}}
\newcommand{\ees}{\end{eqnarray}}
\newcommand{\eens}{\nonumber\end{eqnarray}}

\newcommand{\bra}[1]{\big{\langle}#1\big{|}\,}
\newcommand{\ket}[1]{\,\big{|}#1\big{\rangle}}
\newcommand{\bracket}[2]{\big{\langle}{#1}\big{|}{#2}\big{\rangle}}
\newcommand{\ave}[1]{\big{\langle}#1\big{\rangle}}

\renewcommand{\/}{\over}
\renewcommand{\d}{\partial}
\newcommand{\ddt}{{\d\/\d t}}

\newcommand{\sumi}{\sum_{i=1}^d}
\newcommand{\sumij}{\sum_{i,j=1}^d}
\newcommand{\summu}{\sum_{\mu=0}^d}
\newcommand{\summn}{\sum_{\mu,\nu=0}^d}

\newcommand{\no}[1]{{\,:\kern-0.7mm #1\kern-1.2mm:\,}}
\newcommand{\vect}{{\mathfrak{vect}}}
\newcommand{\map}{{\mathfrak{map}}}
\newcommand{\g}{\mathfrak{g}}

\newcommand{\one}{{(1)}}
\newcommand{\two}{{(2)}}
\newcommand{\nnn}{{(n)}}
\newcommand{\llb}{{(\ell)}}

\renewcommand{\one}{{1}}
\renewcommand{\two}{{2}}
\renewcommand{\nnn}{{n}}
\renewcommand{\llb}{{\ell}}

\newcommand{\muhat}{\hat\mu}
\newcommand{\nuhat}{\hat\nu}
\newcommand{\ihat}{\hat\imath}
\newcommand{\jhat}{\hat\jmath}

\newcommand{\w} {\omega}
\newcommand{\dlt} {\delta}
\newcommand{\eps} {\epsilon}
\newcommand{\si} {\sigma}
\newcommand{\al} {\alpha}
\newcommand{\bt} {\beta}
\newcommand{\ka} {\kappa}
\newcommand{\la} {\lambda}

\newcommand{\mn}{{\mu\nu}}
\newcommand{\ij}{{ij}}
\newcommand{\ab}{{\alpha\beta}}

\newcommand{\xx}{{\mathbf x}}
\newcommand{\yy}{{\mathbf y}}
\newcommand{\qq}{{\mathbf q}}
\newcommand{\pp}{{\mathbf p}}
\newcommand{\PP}{{\mathbf P}}
\renewcommand{\AA}{{\mathbf A}}
\newcommand{\BB}{{\mathbf B}}
\newcommand{\EE}{{\mathbf E}}
\newcommand{\uu}{{\mathbf u}}
\newcommand{\vv}{{\mathbf v}}
\newcommand{\kk}{{\mathbf k}}
\newcommand{\kka}{{\mathbf \kappa}}
\newcommand{\mm}{{\mathbf m}}
\newcommand{\nn}{{\mathbf n}}
\newcommand{\rr}{{\mathbf r}}
\renewcommand{\ss}{{\mathbf s}}

\newcommand{\dx}{d^{d+1}\!x}
\newcommand{\dy}{d^{d+1}\!y}
\newcommand{\dk}{d^{d+1}\!k}

\newcommand{\dxx}{d^d\!x}
\newcommand{\dyy}{d^d\!y}
\newcommand{\dkk}{d^d\!k}
\newcommand{\dkka}{d^d\!\kappa}

\newcommand{\dfx}{d^4\!x}
\newcommand{\dfk}{d^4\!k}
\newcommand{\dEx}{d^3\!x}
\newcommand{\dEk}{d^3\!k}

\newcommand{\e}{{\mathrm e}}
\newcommand{\half}{{1\/2}}

\newcommand{\noll}{{\hat 0}}
\newcommand{\zero}{{\mathbf 0}}
\newcommand{\dzero}{\dlt^{d-1}(\zero)}

\newcommand{\onetwo}{1 \leftrightarrow 2}

\renewcommand{\L}{{\mathcal L}}
\newcommand{\J}{{\mathcal J}}
\newcommand{\HH}{{\mathcal H}}

\newcommand{\PB}{{PB}}
\newcommand{\MpP}{\sqrt{M^2+(\pp - \PP)^2}}
\newcommand{\MpPA}{\sqrt{ M^2 + (\pp - \PP + e\AA(\zero))^2 }}
\newcommand{\stwk}{\sqrt{2|\kk|}}

\newcommand{\ketku}{\ket{{\{\kk\}; \uu}}}
\newcommand{\ketkx}{\ket{{\{\kk\}; \xx}}}

\newcommand{\RR}{{\mathbb R}}
\newcommand{\CC}{{\mathbb C}}
\newcommand{\ZZ}{{\mathbb Z}}
\newcommand{\NN}{{\mathbb N}}

\title{{The physical observer I: Absolute and relative fields}}

\author{T. A. Larsson \\
Vanadisv\"agen 29, S-113 23 Stockholm, Sweden\\
email: thomas.larsson@hdd.se}

\maketitle
\begin{abstract}
Quantum Jet Theory (QJT) is a deformation of QFT where also the quantum
dynamics of the observer is taken into account. This is achieved by
introducing relative fields, labelled by locations measured by rods
relative to the observer's position. In the Hamiltonian formalism, the
observer's momentum is modified: $p_i \to p_i - P_i$, where $P_i$ is the
momentum carried by the field quanta. The free scalar field, free
electromagnetism and gravity are treated as examples. Standard QFT results
are recovered in the limit that the observer's mass $M \to \infty$ and its
charge $e \to 0$. This limit is well defined except for gravity, because
$e = M$ in that case (heavy mass equals inert mass). In a companion paper
we describe how QJT also leads to new observer-dependent gauge and diff
anomalies, which can not be formulated within QFT proper.
\end{abstract}

\vskip 3cm

\section{Physical motivation}

Every experiment is an interaction between a system and a detector,
and the result depends on the physical properties of both. We want
to eliminate the detector dependence as far as possible, but not
further than that. The main thesis in this paper is that the
detector's mass can not be eliminated in the presence of gravity.

Every physical detector\footnote{We will use the words ``detector'' and
``observer'' synonymously. This does of course not imply that the
observer is human or even animate.} has some charge $e$ (because
otherwise it can not interact with the system and thus not observe
it) and some finite and nonzero mass $M$. To extract the
detector-independent physics, we want to take the joint limit $e \to 0$
(so the detector does not perturb the fields) and 
$M \to \infty$ (so the detector follows a well-defined, classical path
in spacetime). If $M \neq \infty$, the detector's position and
velocity at the same instant do not commute, and hence its worldline
is subject to quantum fluctuations.

The joint limit $e \to 0$, $M \to \infty$ is described by QFT. This
limit is well defined for non-gravitational interactions, but runs
into serious trouble when gravity is taken into account. The reason
is that the gravitational charge is closely related to mass -- the
heavy mass equals the inert mass. This observation immediately leads
to a successful postdiction: QFT is incompatible with gravity.

To remedy this problem, we propose to replace QFT by a more complete
theory, which explicitly describes both the quantum-mechanical detector 
and the quantized fields. This theory is, or will be, called Quantum 
Jet Theory (QJT). The strategy for doing this is to introduce the 
detectors's quantized worldline, defined operationally by the readings 
of clocks and rods. All fields are expanded a Taylor series around the
detector's worldline, and physics is formulated in terms of the
Taylor coefficients rather than in terms of the fields themselves.
The motivates the name QJT, because in mathematics a jet is
essentially the same thing as a Taylor expansion\footnote{More
precisely, a $p$-jet is an equivalence class of functions; two
functions belong to the same class if all derivatives up to order
$p$, evaluated at some point $q$, are the same. Since a $p$-jet has
a unique representative which is a polynomial of order at most $p$,
namely the truncated Taylor series around $q$, we may canonically
identify jets and Taylor expansions.}.

Historically, QJT grew out of the projective representation theory
of the algebra of spacetime diffeomorphisms, i.e.
the multi-dimensional Virasoro algebra. In particular, it was noted
in \cite{Lar98} that off-shell representations of lowest-energy type
must be formulated in terms of trajectories of jets rather than in
terms of the spacetime fields themselves. Several flawed attempts to
apply this insight to physics were made, cf. \cite{Lar04,Lar06a}. In
\cite{Lar07}, the correct treatment was finally found, at least for
the harmonic oscillator; the term QJT was also coined in that paper.

Unfortunately, the formalism developed in those papers was quite unwieldy,
for two reasons. First, working with jets rather than fields is
complicated by the fact that jets do not mix nicely with nonlocal
integrals such as the action functional or the Hamiltonian. Second, I
attempted to do canonical quantization in a manifestly covariant way,
identifying phase space with the space of histories. It is the purpose of
the present paper to simplify the formulation of QJT by avoiding these
complications. We break manifest covariance, and work with the fields 
themselves regarded as generating functions for their Taylor series. This
makes it far simpler to extract the physical content of QJT.

In a companion paper \cite{Lar08b}, we discuss the appearance of gauge
and diff anomalies. Contrary to QFT, gauge anomalies are a generic
feature in QJT. The relevant cocycles expicitly depend on the observer's
trajectory, and can hence not be formulated within QFT, where the
observer is never introduced. Because these new gauge anomalies 
have an unconventional form (e.g., for Yang-Mills theory they are 
proportional to the second Casimir rather than to the third), standand
intuition about gauge anomalies does not apply; in particular, these
gauge anomalies do not necessarily render the theory inconsistent.

\section{Partial and complete observables in QFT and QJT}

Recall the distinction between partial and complete variables made by
Rovelli \cite{RovPart}:
\begin{itemize}
\item
{\em Partial observable}: a physical quantity to which we can
associate a (measuring) procedure leading to a number.
\item
{\em Complete observable}: a quantity whose value can be predicted
by the theory (in classical theory); or whose
probability distribution can be predicted by the
theory (in quantum theory).
\end{itemize}

A physical experiment always consists of at least two measurements:
the reading of the detector and the reading of a clock. The partial
observables are thus $\{A, t\}$, where $A$ is the quantity of
interest and $t$ is time. Although both partial observables can be
measured, neither can be predicted without knowledge about the
other. What can be predicted is the complete observable $A(t)$, the
value of $A$ at time $t$. Of course, we can only make predictions,
or check that our assumed dynamics is correct, provided that we know
the state of the system. Hence we must first measure the partial
observables $\{A, t\}$ sufficiently many times to determine the
state. Once that is done, the outcome of further observations is
predicted by the theory.

Complete observables correspond to self-adjoint operators in quantum
mechanics, partial observables do not. The complete observable $A(t)$
is the value of $A$ at time $t$; since the measurement of this value
is subject to quantum flucuations, it is described by an operator.
In contrast, time $t$ itself is a partial observable which serves to
localize the experiment in time, and as such it is a c-number
parameter. Perhaps the distinction is made clearest by the questions
answered by the different types of observables: $A(t)$ answers "What
is the value of $A$ at time $t$?" and $t$ (or $t(t)$) answers "What
is the value of time at time $t$?". Clearly, the answer to the
second question has no room for quantum fluctuations, and hence it
is given by a c-number parameter rather than an operator. In my
opinion, this settles the apparent paradox with Pauli's theorem
\cite{Pau26}, which asserts that there can be no time operator in
quantum mechanics (provided that the energy is bounded from below).

More symmetrically, a complete observable is a correlation $(A, t)$
between partial observables. If the relation between $A$ and $t$ is
monotonous, we can regard this complete observable either as the
value $A(t)$ of $A$ at time $t$, or the value $t(A)$ of $t$ at
detector reading $A$. Either way the complete observable is subject
to quantum fluctuations and thus given by an operator. The monotony
assumption means that $A$ is another clock. E.g., $t$ could be the
observer's proper time $\tau$ (the ticks of the local clock), whereas 
$A$ could be reference time broadcasted from a GPS (Global Positioning
System) satellite. If the experiment described by $A$ amounts to the 
detection of one of these broadcasted signals, it may be viewed as 
another definition of time, and presumably a more accurate one 
than the reading of a local clock device. The observed value of 
broadcasted time at a given observed value of the local clock is clearly 
subject to quantum fluctuations, and hence described by an operator.

Let us now turn to field theory. In QFT the complete observables are
fields $\phi(x)$, where $x = x^\mu$ is a spacetime point. In any 
experiment, we hence measure two kinds of partial observables:
\begin{itemize}
\item[$\phi$:\ ]
The value of the field, measured by our experimental apparatus.
\item[$x^\mu$:\ ]
The detector's spacetime location, measured by rods and clocks.
\end{itemize}
Rather than using rods and clocks, $x^\mu$ can more conveniently be
measured using GPS receivers; we will therefore refer to $x^0$ as
{\em GPS time} and $x^i$ as {\em GPS position} \cite{RovGPS}.

However, there are subtle physical problems with using the complete
observables $(\phi, x)$. The first problem is that we need to know
the state of the system in order to make predictions, and infinitely
many observations are required to determine the state uniquely.
Typically, we must determine the values of the field throughout an
equal-time surface, say $x^0 = 0$. Rovelli suggests that one should
avoid this problem by making additional assumptions about the state
\cite{RovPart}, something which I find unattractive.

A second problem is that a single detector can only measure the
field at a single point on a simultaneity surface, namely where
its worldline intersects the surface. Hence we need an array of
detectors, each equipped with a separate GPS receiver. However, at
time $t=0$ the master detector can not know about the full state
at this time; only at some later time $t=T$, when the information from
the most distant slave detector has reached the master, can the full
state back at $t = 0$ be known. Moreover, to determine
the identity of an individual detector, we need to measure a new
partial observable $n$, which can not be specified to arbitrary
precision if the spacetime location $x$ is a c-number. Starting from
the partial observables $\{n, x, \phi\}$, we can may take $x$ as the
independent variable, hence a c-number. Then the complete observable
$n(x)$, which tells us which one of the detectors is located at
$x$, is an operator. Alternatively, we can ask about the measurement
in detector $n$, but then the complete observable $x(n)$, the
location of this particular detector, becomes an operator. Thus the
spacetime location and the detector's identity can not be
simultaneously specified to arbitrary precision.

Finally, since the detector's position $q^i$ is a partial observable,
it can only be measured but not predicted. This is obviously incorrect
for physical detectors, which move according to some equations of motion.
{F}rom a physical point of view, the difference between QJT and QFT is
that the former takes the detector's nontrivial quantum dynamics into
account\footnote{This difference can be illustrated by the following
exaggerated example. Consider an experiment at the LHC. We can measure
the detector's location (close to Geneva), its velocity (close to zero),
and time (year 2008). These are partial observables which serve to
localize the fields, and within QFT nothing can be said about the
detector's location in the future. In contrast, QJT also describes the
detector's dynamics, and hence the detector's position $q^i(t)$ is a
complete observable whose future values can be predicted (it is likely to
remain close to Geneva).}.

In QJT we have three types of partial observables: 
\begin{itemize}
\item[$\phi$:\ ]
The value of the field, measured by our experimental apparatus.
\item[$q^\mu$:\ ]
The detector's spacetime position, measured by its GPS receiver.
As the notation indicates, $q^\mu$ is assumed to transform as a 
spacetime vector.
\item[$\tau$:\ ]
The detector's proper time, measured by a local clock.
\end{itemize}
Note that we have two time
observables, proper time $\tau$ and GPS time $q^0$. {F}rom this set of
partial observables, we can construct two types of complete
observables: $\phi(\tau)$, the reading of the detector when proper time
is $\tau$, and $q^\mu(\tau)$, the reading of the GPS receiver when proper
time is $\tau$. Unlike the situation in QFT, these observables can be
measured by a single detector, so we have no problems with
nonlocality.

There is of course nothing special about proper time. We can (and
eventually will) instead use GPS time $q^0$ as our independent time
variable, which makes proper time $\tau(q^0)$ into a complete observable.
To make the treatment more symmetrical, we introduce an arbitrary 
timelike parameter $t$ as our independent variable. The QJT observables
then become functions of $t$: $q^\mu(t)$, $\tau(t)$ and $\phi(t)$.
The timelike parameter $t$ is not physical since the theory now has a
gauge symmetry of reparametrizations of the observer's trajectory; the
reparametrization generators $L(t)$ obey the Witt algebra
\be
[L(t), L(t')] &=& (L(t) + L(t')) \dot\dlt(t-t')
\label{repar}
\ee

The QJT observables considered so far evidently contain much less
information than QFT observables, because they only know about the field 
along the detector's worldline. In particular, we can not predict 
anything, because every field theory prediction involves partial 
derivatives transverse to the observer's 
trajectory\footnote{The obvious exception is when the number of spatial
dimensions $d=0$, i.e. quantum mechanics. Like QFT, QJT reduces to 
quantum mechanics in this case, because the observer's location can not
fluctuate when space consists of a single point.}.
Fortunately this problem can easily be solved by measuring further
local data. The most general complete observable that a local
detector can measure at time $t$ is not just the field $\phi(t)$
itself, but also the gradient 
$\phi_{,\mu}(t) = \d_\mu \phi(x)|_{x = q(t)}$, as well as higher
partial derivatives of the field such as $\phi_{,\mu\nu}(t)$.
Clearly, the new observables constructed in this way are not all
independent. There are constraints relating derivatives in the
temporal direction to time evolution, the simplest one being
\be
\dot \phi(t) = \dot q^\mu(t) \phi_{,\mu}(t).
\label{Dot1}
\ee
We can use this equation to eliminate one component of the gradient. If 
we use the reparametrization symmetry (\ref{repar}) to fix $q^0(t) = t$,
all partial derivatives with at 
least one $\mu=0$ index can be eliminated by the relations
\be
\phi_{,0\mu_1..\mu_p}(t) = \dot \phi_{,\mu_1..\mu_p}(t)
- \dot q^i(t) \phi_{,i\mu_1..\mu_p}(t),
\ee
which are analogous to (\ref{Dot1}). The QJT observables 
$\phi_{,\mu_1..\mu_p}(t)$ contain the same amount information as the QFT
observables $\phi(x)$ in a neighborhood of the observer's trajectory
$q^\mu(t)$.

Let us summarize the main differences between QFT and QJT:
\begin{itemize}
\item
In QFT, the partial observables are $\{x^\mu,\phi\}$, and the
complete observables $\phi(x)$ answer the question: "What does the
detector measure when the GPS receiver measures $x^\mu$?". The
answer requires an array of detectors, whose identity and position
can not both be measured sharply, and whose time evolution can not
be predicted.
\item
In QJT, the partial observables are $\{t, \tau, q^\mu, \phi, \phi_{,\mu},
\phi_{,\mu\nu}, ...\}$. The complete observables  $\tau(t)$, $q^\mu(t)$,
$\phi(t)$, $\phi_{,\mu}(t)$, ... answer the questions: "What do the
local clock, the GPS receiver and the field detector measure when the 
time parameter is $t$?". This can be answered by a single, local detector,
whose position evolves in time in a predictable manner.
\end{itemize}

\section{Absolute and relative fields}
\subsection{ Spacetime fields}

We now start with the formalization of the physical discussion in the
previous section. Consider some field $\phi(x)$ over $(d+1)$-dimensional
spacetime, where $x = (x^\mu) \in \RR^{d+1}$ are spacetime coordinates. 
In QFT, these coordinates are measured relative some absolute, fixed
origin\footnote{Physically, the fixed origin may be thought of as the 
location of an infinitely heavy observer, e.g. a GPS satellite.}. To 
emphasize this point, we call this an {\em absolute field} $\phi_A(x)$. 
In QJT we instead consider the {\em relative field} $\phi_R(x,t)$, 
where the spacetime coordinates are measured relative to the physical 
observer's spacetime location $q^\mu(t)$.
The time coordinate $t$ will soon be identified with GPS time: 
$t = x^0 = q^0(t)$. The important difference is that $q^i(t)$ is 
``on the other side of the rod'', i.e. that positions are
measured relative to the observer's location rather than relative to 
some fixed origin. The observer's position at time $t$ can be predicted
once we know the observer's quantum dynamics, and hence $q^i(t)$ must
be a complete observable, which becomes an operator after quantization.
This is the essential novelty in QJT.

Absolute and relative fields are related by
\bes
\phi_R(x,t) &=& \phi_A(x+q(t)), 
\nlb{AR}
\phi_A(x) &=& \phi_R(x-q(t),t),
\eens
because absolute and relative position are related by $x_A = x_R + q(t)$.
As the notation indicates, absolute fields do not depend on the 
time parameter $t$. {F}rom $d\phi_A(x)/dt = 0$ it follows that relative
fields satisfy
\be
{\d\/\d t} \phi_R(x,t) - \dot q^\mu(t) \d_\mu \phi_R(x,t) \equiv 0.
\label{dfR}
\ee

In QFT we are used to operator-valued functions like $\phi_A(x)$, which 
depends on c-number arguments, but what is the meaning of
$\phi_R(x,t) = \phi_A(x+q(t))$,
where the argument is itself an operator? To answer this question, we
consider absolute fields being defined by their Taylor series, {\em viz.}
\be
\phi_A(x) = \sum_{m\in\NN^{d+1}} {1\/m!} \phi_{,m}(t) (x-q(t))^m.
\label{Ajet}
\ee
Here we use standard multi-index notation introduced e.g. in 
\cite{Lar04}: $m = (m_0, m_1, ..., m_d) \in \NN^{d+1}$,
$m! = m_0! m_1! ... m_d!$, 
$(x-q)^m = (x^0 - q^0)^{m_0}(x^1 - q^1)^{m_1} ... (x^d - q^d)^{m_d}$.
Denote by $\hat\mu$ a unit vector in the $\mu$:th direction, so that
$m+\muhat = (m_0, ...,m_\mu+1, ..., m_d)$. The Taylor
cooefficients
\be
\phi_{,m}(t) = \d_m\phi_A(q(t),t)
= \underbrace{\d_0 .. \d_0}_{m_0} \underbrace{\d_1 .. \d_1}_{m_1} .. 
\underbrace{\d_d .. \d_d}_{m_d} \phi_A(q(t),t)
\label{jetdef}
\ee
can be identified with the $|m|$:th order derivative of $\phi_A(x,t)$,
evaluated on the observer's trajectory $q^\mu(t)$. Note the difference
between $m = 0$ and $m = \hat0$:
$\phi_{,0}(t) = \phi_A(q(t))$ but $\phi_{,\hat0}(t) = \d_0\phi_A(q(t))$.

Analogously, the corresponding relative field is defined as the 
MacLaurin series
\be
\phi_R(x) = \sum_{m\in\NN^{d+1}} {1\/m!} \phi_{,m}(t) x^m.
\label{Rjet}
\ee
The identity (\ref{dfR}) becomes
\be
\dot \phi_{,m}(t) - \sum_{\mu=0}^d \dot q^\mu(t) \phi_{,m+\hat\mu}(t) = 0,
\label{dfm}
\ee
which contains (\ref{Dot1}) as a special case.
Although we will not work explicitly with the series (\ref{Ajet}) 
and (\ref{Rjet}) in this paper, they are useful as an unambiguous 
definition of fields with operator-valued arguments. All formulas for
the fields can readily be transformed into hierarchies of equations
for the Taylor coefficients in (\ref{Ajet}) or (\ref{Rjet}). This is
the motivation for the name QJT (Quantum Jet Theory).

The absolute field $\phi_A(x)$ depends on two types of coordinates:
\begin{itemize}
\item[$x^0$]
GPS time relative to a fixed origin. This is a partial observable 
and hence a c-number parameter.
\item[$x^i$]
GPS position relative to a fixed origin. This is a partial 
observable and hence a c-number parameter.
\end{itemize}
These coordinates can be combined into a spacetime vector 
$x^\mu = (x^0, x^i)$.

In contrast, we can consider no less than six different spacetime
coordinates for the relative field $\phi_R(x,t)$:
\begin{itemize}
\item[$t$]
An arbitrary timelike gauge parameter, which can be eliminated by 
gauge-fixing the reparametrization symmetry (\ref{repar}).
\item[$q^0(t)$]
The observer's GPS time relative to a fixed origin, at parameter time 
$t$. This is a complete observable once the observer's dynamics is 
specified, and hence an operator.
\item[$q^i(t)$]
The observer's GPS position relative to a fixed origin, at parameter
time $t$. This is a complete observable once the observer's dynamics 
is specified, and hence an operator.
\item[$x^0$]
GPS time relative to the observer's time $q^0(t)$. This is a partial 
observable and hence a c-number parameter.
\item[$x^i$]
GPS position relative to the observer's position $q^i(t)$. This is a 
partial observable and hence a c-number parameter.
\item[$\tau(t)$]
The observer's proper time, as measured by a local clock, at parameter
time $t$. It depends on the metric $g_\mn$ as 
$\tau(t)^2 = g^A_\mn(q(t)) \dot q^\mu(t) \dot q^\nu(t)$ 
(absolute field) or 
$\tau(t)^2 = g^R_\mn(0) \dot q^\mu(t) \dot q^\nu(t)
= g_{\mn,0}(t) \dot q^\mu(t) \dot q^\nu(t)$
(relative field).
\end{itemize}
These coordinates are combined into two spacetime vectors
$q^\mu(t) = (q^0(t), q^i(t))$ and $x^\mu = (x^0, x^i)$, whereas proper
time $\tau$ is a Lorentz scalar.

\subsection{ Space-time decomposition }

By definition (\ref{AR}), the relative field $\phi_R(x,t)$ depends on
three time variables $x^0$, $t$ and $q^0(t)$. This is clearly 
two too many, and this will lead to various complications, e.g. the 
well-known type of gauge symmetry associated with parametrized time.
Since we do not wish to deal with this type of complication here, 
but rather want to extract the physical consequences of relative 
fields, we eliminate the two extra time variables.

First use the reparametrization freedom to equal the time parameter to 
the detector's GPS time,
\be
q^0(t) = t.
\label{q0=t}
\ee
Moreover, we foliate spacetime into slices of constant GPS time,
\be
x^0 = t.
\label{x0=t}
\ee
The absolute field (\ref{Ajet}) then takes the form
\be
\phi_A(t,\xx) = \sum_{\mm\in\NN^d} 
 {1\/\mm!} \phi_{,\mm}(t) (\xx-\qq(t))^\mm,
\label{Aj}
\ee
where boldface denotes $d$-dimensional spatial vectors, e.g.
$\xx = (x^i)$, $\mm = (m_i)$. As usual, greek indices $\mu$, $\nu$ run
over spacetime directions and latin indices $i$, $j$ label space 
directions. In (\ref{Aj}) there is a single
time coordinate $t$. The pair $(t,\xx)$ is enough to uniquely label
the spacetime point where the field $\phi_A$ is measured, and hence they
are partial, c-number observables.
The relative time coordinate becomes $x^0 = 0$, and the relative 
field (\ref{Rjet}) becomes
\be
\phi_R(t,\xx) = \sum_{\mm\in\NN^d} {1\/\mm!} \phi_{,\mm}(t) \xx^\mm.
\label{Rj}
\ee

Both conditions (\ref{q0=t}) and (\ref{x0=t}) break manifest Lorentz
symmetry. The latter is just the ordinary foliation in the
Hamiltonian formalism, and hence it does not break true Lorentz 
invariance. In contrast, the former condition has no counterpart in QFT, 
and it is possible that it also breaks true Lorentz symmetry. This is not
unphysical, because in QJT there is a distinguished direction in
spacetime, namely parallel to the physical observer's trajectory.

In general-covariant theories it is likely that if manifest
diffeomorphism invariance is broken, so is true diffeomorphism invariance.
The foliation (\ref{x0=t}) is problematic already in QFT, because the
notion of an equal-time surface depends on the quantized metric.
Moreover, the assumption (\ref{q0=t}) was studied in a diffeomorphism
algebra context in
\cite{Lar98}, section 7. Although it does not affect the
spacetime diffeomorphism algebra proper, reparametrization cocycles
associated with the Witt algebra (\ref{repar}) transmute into complicated
diffeomorphism cocycles, which are noncovariant because they single out
the $x^0$ direction.

{F}rom (\ref{dfR}) and $\dot q^0(t) = 1$ we find that
\bes
\d_0\phi_A(t,\xx) &=& \dot\phi_A(t,\xx), 
\label{d0fR} \\
\d_0\phi_R(t,\xx) &=& \dot\phi_R(t,\xx) - \dot q^i(t)\d_i \phi_R(t,\xx),
\eens
where we denote the partial derivative with respect to $t$ by a dot:
\be
\dot \phi_R(t,\xx) \equiv {\d\/\d t}\phi_R(t,\xx).
\ee
The expansions (\ref{Aj}) and (\ref{Rj}) only depend on the spatial 
components $\phi_{,\mm}$, but (\ref{d0fR}) allows us to recursively 
recover the time derivatives by
\be
\phi_{,\mm+\hat0}(t) = \dot \phi_{,\mm}(t) 
- \sum_{i=1}^d \dot q^i(t) \phi_{,\mm+\ihat}(t).
\ee
This notation is self-consistent, because the spacetime decomposition
of the relative field $(\d_0)^n\phi_R(x)$ is
\be
(\d_0)^n\phi_R(t,\xx) = \sum_{\mm\in\NN^d} 
{1\/\mm!} \phi_{,\mm+n\noll}(t) \xx^\mm,
\ee
which is precisely what we get by applying to (\ref{Rj}) the operator 
$\d_0$, defined in (\ref{d0fR}).

\subsection{Poisson brackets}

The configuration space in QJT is spanned by the Taylor coefficients
$\phi_{,\mm}$ and the observer's position $q^i$. Introduce canonical 
momenta $\pi^{,\nn}$ and $p_j$, which by definition satisfy the Poisson
brackets
\bes
[\phi_{,\mm}(t), \pi^{,\nn}(t)]_\PB &=& \dlt^\nn_\mm, \nle
{[}q^i(t), p_j(t)]_\PB &=& -\dlt^i_j.
\eens
All other equal-time brackets are assumed to vanish. In Minkowki 
space, vector indices are raised and lowered by means of the flat metric
$\eta_\ij$, e.g.
\be
q_i(t) = \eta_\ij q^j(t) = - q^i(t).
\ee
In general relativity, we instead use the metric field on the 
observer's trajectory, e.g.
\be
\dot q_i(t) = g^R_\ij(t,\zero) \dot q^j(t) 
= g^A_\ij(t,\qq(t)) \dot q^j(t).
\ee
However, the upper multi-index in $\pi^{,\mm}$ can not be lowered with
the Minkowski metric in a meaningful way. Instead, the natural definition
of $p$-jet momentum with a multi-index downstairs is
\be
\pi_{,\mm}(t) =  (-)^\nn \d_{\mm+\nn}\dlt(\zero) \pi^{,\nn}(t),
\label{pilower}
\ee
where $\dlt(\zero)$ is the delta function evaluated at the origin and
$\d_{\mm+\nn}$ denotes the $(\mm+\nn)$:th derivative, defined as in
(\ref{jetdef}). This is of course a formal expression, which must be
given a definite meaning. E.g., we may define the delta function as the 
limiting value of a family of narrowly peaked Gaussians.

The nonzero Poisson brackets in jet space are
\bes
[\phi_{,\mm}(t), \pi_{,\nn}(t)]_\PB &=& (-)^\nn \d_{\mm+\nn}\dlt(\zero), 
\nlb{jetCCR}
[q^i(t), p_j(t)]_\PB &=& \dlt^i_j.
\eens
We can now define the absolute and relative field momenta by
\bes
\pi_A(t,\xx) &=& \sum_{\mm\in\NN^d}
{1\/\mm!} \pi_{,\mm}(t) (\xx-\qq(t))^\mm, \nle
\pi_R(t,\xx) &=& \sum_{\mm\in\NN^d} 
{1\/\mm!} \pi_{,\mm}(t) \xx^\mm.
\eens
The absolute fields satisfy the nonzero Poisson brackets
\bes
[\phi_A(t,\xx), \pi_A(t,\xx')]_\PB &=& \dlt(\xx-\xx'), \nl
{[}q^i(t), p_j(t)]_\PB &=& \dlt^i_j, \nle
{[}p_i(t), \phi_A(t,\xx)]_\PB &=& \d_i \phi(t,\xx), \nl
{[}p_i(t), \pi_A(t,\xx)]_\PB &=& \d_i \pi(t,\xx).
\eens
The absolute field and its momentum do not commute with the observer's
momentum, because
\be
[p_i(t), (x - q(t))^\mm]_\PB = m_i (x - q(t))^{\mm - \ihat}.
\ee
In contrast, the relative field is defined by the MacLaurin series
(\ref{Rj}) and is independent of $q^i(t)$. It satisfies the
Heisenberg algebra with nonzero brackets
\bes
[\phi_R(t,\xx), \pi_R(t,\xx')]_\PB &=& \dlt(\xx-\xx'), 
\nlb{Poisson_R}
[q^i(t), p_j(t)]_\PB &=& \dlt^i_j.
\eens
In particular, 
\be
[p_i(t), \phi_R(t,\xx)]_\PB = [p_i(t), \pi_R(t,\xx)]_\PB = 0.
\ee

We now see why the jet momentum with lower multi-index must be defined
as in (\ref{pilower}). Computing the $[\phi_A(\xx),\pi_A(\xx')]$
bracket using their Taylor series defintion, we find
\bes
\sum_\mm \sum_\nn {1\/\mm!} {1\/\nn!} (-)^\nn 
\d_{\mm+\nn}\dlt(\zero) (\xx-\qq)^\mm (\xx'-\qq)^\nn &=& \\
=\ \sum_\rr {1\/\rr!} \d_\rr\dlt(\zero) (\xx-\xx')^\rr
&=& \dlt(\xx-\xx'),
\eens
where the intermediate expression is the expansion of the delta 
function around the origin.

\subsection{ Dynamics }

Both the fields and the observer's trajectory are dynamical degrees of
freedom in QJT, and hence we need to introduce dynamics for both. The
field part of the action is the same as in field theory, but we must also
add terms describing the observer's dynamics and the field-observer
interaction. The observer is assumed to be a point particle travelling
along the trajectory $q^\mu(t)$, in accordance with the definition of
absolute and relative fields in (\ref{AR}). One could in principle
consider extended observers, but an irreducible observer is pointlike.

Consider a general field theory with several absolute fields 
$\phi^a_A(x)$, also labelled by another index $a$.
We posit that the action is of the form $S = S_\phi + S_q$, where
\bes
S_\phi &=& \int \dx\, \L_\phi(\phi_A, \d_\mu\phi_A) \nl
&=& \iint dt\,\dxx\, \L_\phi(\phi_A, \d_0\phi_A, \d_i\phi_A), 
\label{SgenAbs}\\
S_q &=& \int dt\, L_q(q, \dot q, \phi_A(\qq(t))),
\eens
where we recall from (\ref{d0fR}) that $\d_0\phi^a_A = \dot \phi^a_A$.
The Lagrangian is thus of the form 
\be
L(t) = L_\phi(t) + L_q(t) = \int \dxx\, \L_\phi(t,\xx) + L_q(t).
\label{Lag}
\ee
The first part has the standard field theory form, and we assume that
the dynamics of the observer and the observer-field interaction is 
described by a term of the form $S_q$.

In the field part of the action, the integrand is a function of 
$\phi_A(t,\xx)$ and its derivatives. Such an integral can be rewritten as
\bes
S_\phi &=& \iint dt\,\dxx\, F(\phi_A(t,\xx), \d_\mu \phi_A(t,\xx)) \nl
&=& \iint dt\,\dxx\, F(\phi_R(t,\xx-\qq), \d_\mu \phi_R(t,\xx-\qq)) \\
&=& \iint dt\,\dyy\, F(\phi_R(t,\yy), \d_\mu \phi_R(t,\yy)),
\eens
where $\yy = \xx - \qq$, and we assume that we are free to make a linear
shift in the measure, i.e. $\dxx = \dyy$. Classically, such a shift can
be done, at least as long as boundary conditions are ignored. Whether
this assumption is as innocent on the quantum level is less clear, since 
the difference between $\dxx$ and $\dyy$ is
an operator $d^dq(t)$. On the other hand, since the integrals over $x$
and $y$ are equivalent classically, it is not obvious which is right
choice after quantization; even if the these integrals would disagree, 
the $y$ integral may well be the physically correct choice. This 
subtlety is ignored in the rest of this paper.

When expressed in terms of relative fields, the action becomes
\bes
S_\phi &=& \iint dt\,\dxx\, \L_\phi(\phi_R, \d_0\phi_R, \d_i\phi_R), 
\nlb{SgenRel}
S_q &=& \int dt\, L_q(q, \dot q, \phi_R(\zero)),
\eens
where we recall from (\ref{d0fR}) that 
\be
\d_0\phi^a_R = \dot \phi^a_R - \dot q^i \d_i\phi^a_R.
\ee
The field part $S_\phi$ is thus assumed to be independent of the
observer's location, except implicitly through the definition of $\d_0$, 
and the observer-field interaction is encoded in $S_q$.
The action (\ref{SgenRel}) leads to the canonical momenta
\bes
\pi^R_a(\xx) &=& {\d\L_\phi\/\d\dot\phi_R^a(\xx)}
\ =\ {\d\L_\phi\/\d\d_0\phi_R^a(\xx)}, 
\label{pqq}\\
p_i &=& {\d L_q\/\d\dot q^i} - 
\int \dxx\, {\d\L_\phi\/\d\d_0\phi_R^a(\xx)} \d_i\phi_R^a(\xx) 
\ =\ {\d L_q\/\d\dot q^i} - P_i,
\eens
where
\bes
P_i &\equiv& \int \dxx\, 
{\d\L_\phi\/\d\d_0\phi_R^a(\xx)} \d_i\phi_R^a(\xx) 
\\
&=& \int \dxx\, \pi^R_a(\xx) \d_i\phi_R^a(\xx)
\ =\ -\int \dxx\, \phi_R^a(\xx)\d_i\pi^R_a(\xx).
\eens
The Hamiltonian is of form
\bes
H &=& \int \dxx\, \pi^R_a(\xx) \dot\phi_R^a(\xx) + p_i \dot q^i
- \int\dxx\, \L_\phi - L_q 
\nlb{HamGen}
&=& H_\phi + H_q,
\eens
where the field part $H_\phi(\phi_R,\pi_R)$
has the same functional form as for absolute fields. To find an explicit
expression for the observer part, we assume that the equation
$p_i = {\d L_q\/\d\dot q^i}(\qq,\dot \qq)$ can be inverted to yield
$\dot q^i = v^i(\qq,\pp)$. Equation (\ref{pqq}) then implies that
\be
\dot q^i = v^i(\qq, \pp + \PP).
\ee
Putting this expression back into (\ref{HamGen}) then yields
\bes
H_q &=& \dot q^i (p_i + P_i) - L_q \nle
&=& (p_i+P_i) v^i(\qq, \pp+\PP) - L_q(\qq, \vv(\qq,\pp+\PP)).
\eens
Comparing this to the corresponding analysis for absolute fields,
which yield $H_q = p_i v^i(\qq,\pp) - L_q(\qq,\vv(\qq,\pp))$,
we see that the passage from absolute to relative fields amounts to
the substitutions
\bes
H_\phi(\phi_A, \pi_A) &\to& H_\phi(\phi_R,\pi_R), 
\nlb{Hfq}
H_q(\qq,\pp,\phi_A(\qq)) &\to& H_q(\qq, \pp+\PP, \phi_R(\zero)).
\eens
To summarize:

\begin{center}
\fbox{ \parbox{0.9\linewidth}
{ \begin{center} 
The passage from absolute to relative fields is equivalent to the 
substitution $p_i \to p_i + P_i$ in the observer part of the Hamiltonian.
\end{center} }
}
\end{center}

The relative Hamiltonian (\ref{Hfq}) leads to the following Hamilton's
equation:
\bes
\dot\phi^a_R(\xx) &=& {\dlt H_\phi\/\dlt\pi^R_a(\xx)} + 
{\d H_q\/\d p_i} \d_i \phi^a_R(\xx), \nl
\dot\pi^R_a(\xx) &=& -{\dlt H_\phi\/\dlt\phi^a_R(\xx)} + 
{\d H_q\/\d p_i} \d_i \pi^R_a(\xx)
- {\dlt H_q\/\dlt \phi^a_R(\xx)} \dlt(\xx), \nl
\dot q^i &=& {\d H_q\/\d p_i}, \\
\dot p_i &=& -{\d H_q\/\d q^i}.
\eens
Combining the definition of $\d_0$ in (\ref{d0fR}) and the evolution
equation for $q^i$, we get
\be
\d_0\phi^a_R(\xx) = \dot \phi^a_R(\xx) 
- {\d H_q\/\d p_i} \d_i\phi^a_R(\xx),
\ee
etc. The first Hamilton's equations can thus be written in the form
\bes
\d_0\phi^a_R(\xx) &=& {\dlt H_\phi\/\dlt\pi_a^R(\xx)}, \\
\d_0\pi_a^R(\xx) &=& -{\dlt H_\phi\/\dlt\phi^a_R(\xx)}
- {\dlt H_q\/\dlt \phi^a_R(\xx)} \dlt(\xx).
\eens
Apart from the last term, which encodes the interaction between the
fields and the observer, this is of the familiar form.

\subsection{Quantization}

A model with relative fields can be canonically quantized as usual.
Replace the Poisson brackets (\ref{Poisson_R}) with commutators
and represent the Heisenberg algebra on a Hilbert space. In the very
simple case that the observer does not interact with the fields, the
Hilbert space becomes the tensor product
$\HH_{tot} = \HH_{field} \otimes \HH_{obs}$. Let $\ket\kk$ and 
$\ket \uu$ be eigenstates of $H_\phi$ and $H_q$, respectively, with
eigenvalues $E_q(\kk)$ and $E_q(\pp(\uu))$, respectively. Moreover, 
assume for simplicity that $\ket\uu$ and $\ket\kk$ are eigenstates of 
$p_i$ and $P_i$, respectively, {\em viz.}
\be
p_i \ket\uu = p_i(\uu) \ket\uu, \qquad
P_i \ket\kk = k_i \ket\kk.
\ee
Under these assumptions, the product state $\ket\kk \otimes \ket\uu$ 
is an eigenstate of the total Hamiltonian, and the eigenvalue is
\be
E(\kk,\uu) = E_\phi(\kk) + E_q(\pp(\uu) + \kk)).
\ee
We deal with quantization in more detail in the examples below.

\subsection{ Observer versus frame dependence }
\label{ssec:frame}

It should be emphasized that observer dependence does not mean that we
work in the observer's rest frame. To the contrary, since we label
spacetime points by their GPS coordinates $x^\mu$, we work in the frame
of the GPS satellites. That we do not work in the observer's rest frame
is easy to see, because the observer's velocity $\dot q^i(t) = dq^i/dq^0
\neq 0$. E.g., if the observer Hamiltonian with absolute fields is $H_ q
= \sqrt{M^2+\pp^2}$, the analogous relative quantity is $H_q =
\sqrt{M^2+(\pp+\PP)^2}$, and not $H_q = M$ which would be the case in 
the observer's rest frame.

\section{ Free scalar field }

\subsection{Action}

The action for a self-interacting scalar field reads 
$S = S_\phi + S_q$, where
\bes
S_\phi &=& \int \dx\, \big( {1\/2} \d_\mu\phi_A \d^\mu \phi_A 
- V(\phi_A) \big), 
\nlb{SphiA}
S_q &=& - M \int dt\, \sqrt{\dot q^\mu \dot q_\mu}.
\eens
In particular, if the scalar field is free and has mass $\w$, the 
potential is $V(\phi) = 1/2\,\w^2 \phi^2$; we denote the mass by $\w$ 
rather than $m$ to avoid confusion with multi-indices. The field part
of the action becomes
\bes
S_\phi = {1\/2} \int \dx\, \big( (\d_0\phi_A)^2 - (\nabla\phi_A)^2
- \w^2 \phi_A^2 \big).
\ees
We now introduce relative fields and eliminate 
reparametrization freedom by (\ref{q0=t}). The action becomes
\bes
S_\phi &=& {1\/2} \iint dt\,\dxx\, \big( (\d_0\phi_R)^2 
- (\nabla\phi_R)^2 - \w^2 \phi_R^2 \big), 
\nlb{Sphi}
S_q &=& - M \int dt\, \sqrt{1 - \dot\qq^2},
\eens
where
\be
\d_0 \phi_R = \dot \phi_R + \dot \qq \cdot \nabla \phi_R
\equiv \dot \phi_R + \dot q_i \d_i \phi_R.
\label{d0fR2}
\ee
Note the change of sign compared to (\ref{d0fR}), due to the contraction
of two lower indices.

\subsection{ Equations of motion }

The Lagrangian is of the form (\ref{Lag}), and 
the Euler-Lagrange equations 
\bes
{\dlt S\/\dlt\phi_R} &=& -{d\/dt}{\d \L_\phi\/\d\dot\phi_R} -
\d_i {\d \L_\phi\/\d(\d_i\phi_R)} + {\d \L_\phi\/\d\phi_R} 
+ {\d L_q\/\d\phi_R}\delta(\xx) = 0, \nl
{\dlt S\/\dlt q_i} &=& 	-{d\/dt}{\d L\/\d\dot q_i} 
+ {\d L\/\d q_i} = 0,
\ees
become
\bes
- \d_0^2 \phi_R(t,\xx) + \nabla^2\phi_R(t,\xx) - \w^2\phi_R(t,\xx)
&=& 0, 
\label{Ephi} \\
{d\/dt}\big( M \gamma(\dot\qq(t)) \dot q_i(t) + P_i(t) \big) &=& 0.
\label{Eq}
\ees
Here we use the standard notation
\be
\gamma(\uu) &\equiv& {1\/\sqrt{1-\uu^2}},
\label{gamma}
\ee
and introduce the operator which measures the field momentum:
\be
P_i(t) &\equiv& \int \dxx\, \d_0\phi_R(t,\xx)\, \d_i \phi_R(t,\xx).
\label{Pi}
\ee
The solution of the Euler-Lagrange equation for $\phi_R$ is 
straightforward. We know that the corresponding absolute field solution 
is a sum over plane waves
\be
\phi_A(t,\xx) = \exp(ik_0t - i\kk\cdot\xx),
\ee
with energy $k_0 = \pm \w_\kk$, where
\be
\w_\kk = \sqrt{\kk^2 + \w^2}.
\label{wk}
\ee
The relative field solution to (\ref{Ephi}) is thus
\be
\phi_R(t,\xx) = \exp(ik_0t - i\kk\cdot(\xx+\qq(t))),
\label{plane}
\ee
with the same dispersion relation. We can now evaluate 
\be
P_j(t) = -k_0 k_j \exp(2ik_0t - 2i\kk\cdot\qq(t)) 
\int \dxx\, exp(-2i\kk\cdot\xx).
\ee
The integral is proportional to $\dlt(\kk)$, and since 
$k_j \dlt(\kk) = 0$, we find that 
\be
P_j(t) = 0. 
\label{Pj=0}
\ee
This result has only been checked for free fields, but it seems likely
to hold generically.
The solution to the evolution equation (\ref{Eq}) is 
\be
M \gamma(\dot\qq(t)) \dot q_j(t) = p_j,
\ee
for some constants $p_j$. Defining velocities $u_j$ by 
$p_j = M\gamma(\uu) u_j$, we finally find that the observer moves along
the straight line
\be
q_j(t) = u_j t + s_j,
\label{straight}
\ee
for some constants $u_j$ and $s_j$.	This is the expected result; since
the field and the observer do not interact, they evolve as a free field
and as a free particle, respectively.

We observed in the previous section that in the Hamiltonian formulation,
the full effect of working with relative fields is to shift the
observer's momentum $\pp \to \pp - \PP$. Equation (\ref{Pj=0}) then
asserts that for classical solutions, $\PP = 0$, so relative fields do
not give anything new classically; QJT only differs significantly from
QFT on the quantum level. This was expected, since the only difference
between QJT and QFT is that the former takes the observer's quantum
dynamics into account.

\subsection {Hamiltonian}

The relative canonical momenta read
\bes
\pi_R(t,\xx) &=& {\d\L_\phi\/\d\dot\phi_R} 
\ =\ \d_0 \phi_R(t,\xx), \nle
p_j(t) &=& {\d L_q\/\d\dot q_j } +  {\d L_\phi\/\d\dot q_j } 
\ =\ M \gamma(\dot\qq(t)) \dot q_j(t) + P_j(t),
\eens
where $\gamma(\dot\qq)$ was defined in (\ref{gamma}), and the definition
of $P_i(t)$ in (\ref{Pi}) becomes
\bes
P_i(t) &=& \int \dxx\, \pi_R(t,\xx)\, \d_i \phi_R(t,\xx) 
\nlb{Pip}
 &=& -\int \dxx\, \d_i\pi_R(t,\xx)\,  \phi_R(t,\xx).
\eens
The Hamiltonian is
\be 
H = \int \dxx\, \pi_R \dot \phi_R + p_j \dot q_j - L 
= H_\phi + H_q,
\ee
where
\bes
H_\phi &=& {1\/2} \int \dxx\, \big( \pi_R^2 + (\nabla\phi_R)^2 
+ \w^2 \phi_R^2 \big), 
\nle
H_q &=& \MpP,
\eens
and $(\pp - \PP)^2 = (p_i - P_i) (p_i - P_i)$. 

Hamilton's equations read
\bes
\dot\phi_R(\xx) &=& {\dlt H\/\dlt\pi(\xx)}
\ =\ \pi(\xx) - {p_i-P_i\/\MpP}\,\d_i\phi_R(\xx), \nl
\dot\pi_R(\xx) &=& -{\dlt H\/\dlt\phi(\xx)}
\ =\ \nabla^2 \phi_R(\xx) - \w^2\phi_R(\xx)
- {p_i-P_i\/\MpP}\,\d_i\pi_R(\xx), \nl
\dot q_i &=& {\d H\/\d p_i}
\ =\ {p_i-P_i\/\MpP}, \\
\dot p_j &=& -{\d H\/\d q_i}\ =\ 0.
\eens
Using (\ref{d0fR2}) in the form
\be
\d_0\phi_R = \dot\phi_R + {(p_i-P_i)\/\MpP}\,\d_i\phi_R,
\ee
Hamilton's equations can be rewritten as
\bes
\d_0\phi_R(\xx) &=& \pi_R(\xx), \nl
\d_0\pi_R(\xx) &=& \nabla^2 \phi_R(\xx) - \w^2\phi_R(\xx), \nl
M\gamma(\dot\qq) \dot q^i &=& p_i - P_i, \nle
\dot p_j &=& 0.
\eens
The solution is of course still given by (\ref{plane}) and 
(\ref{straight}).

\subsection{Quantization}

Now we quantize the theory by replacing Poisson brackets by commutators.
The only nonzero brackets in the Heisenberg algebra are
\bes
[\phi_R(t,\xx), \pi_R(t,\xx')] &=& i\dlt(\xx-\xx'), \nle
[q_i(t), p_j(t)] &=& i\dlt_\ij.
\eens
We use units such that $\hbar=1$, but occationally reinsert factors of
$\hbar$ when needed for clarity. After a Fourier 
transformation in space only, the modes satisfy
\be
[\phi_R(t,\kk), \pi_R(t,\kk')] &=& i\dlt(\kk+\kk').
\ee
Define creation and annihilation operators by
\bes
a_\kk = {1\/\sqrt{2\w_\kk}} 
(\pi_R(\kk) - i\w_\kk \phi_R(\kk)), \nle
a^\dagger_\kk = {1\/\sqrt{2\w_\kk}} 
(\pi_R(\kk) + i\w_\kk \phi_R(\kk)).
\eens
The complete set of nonzero commutation relations is thus
\bes
[a_\kk(t), a^\dagger_{\kk'}(t)] &=& \dlt(\kk + \kk'), \nle
[q_i(t), p_j(t)] &=& i\dlt_\ij,
\eens
where we also remember that the system depends on the observer's
position and momentum.
Introduce the number operator 
\be
N = \int \dkk\, a^\dagger_\kk a_{-\kk}.
\ee
The extra term $P_j$, defined in (\ref{Pip}), takes the form
\be
P_j = i\int \dkk\, k_j \pi_R(-\kk)\phi(\kk) 
= \int \dkk\,  k_j a^\dagger_\kk a_{-\kk}.
\ee
$N$ and $P_j$ satisfy
\bes
[N, a_\kk] = - a_\kk, &\quad&
[N, a^\dagger_\kk] = a^\dagger_\kk, \nle
[P_j, a_\kk] =  k_j a_\kk, &\quad&
[P_j, a^\dagger_\kk] =  k_j a^\dagger_\kk.
\eens
$N$ is thus the number operator for quanta, and $P_j$ the operator that
counts the momentum of the field quanta.

The field part of the Hamiltonian acquires the familiar form
\be
H_\phi = \int \dkk\, \w_\kk a^\dagger_\kk a_{-\kk},
\ee
and the total Hamiltonian is $H = H_\phi + H_q$, where as before
\be
H_q = \MpP.
\ee
We note that $[H_\phi,H_q] = 0$, and that $N$, $p_i$ and $P_i$ commute 
with both $H_\phi$ and $H_q$ and among themselves. We can therefore 
diagonalize $H_\phi$, $H_q$, $N$, $p_i$ and $P_i$ simultaneously. 
The Fock space is spanned by states with $N$ quanta with energy
$H_\phi$ and momentum $P_i$, and the observer has energy $H_q$ and
momentum $p_i$. In particular, the Fock vacuum is a tensor product fully
characterized by the observer's velocity $\uu$:
\be
\ket {0; \uu} = \ket0 \otimes \ket \uu.
\ee
The general $n$-quanta state,
\be
\ketku \equiv \ket{{\kk_\one, ..., \kk_\nnn; \uu}}
= a^\dagger_{\kk_\one} ... a^\dagger_{\kk_\nnn}
\ket 0 \otimes \ket \uu,
\ee
is an unnormalized eigenstate with the following eigenvalues:
\bes
N \ketku &=& n \ketku , \nl
P_j \ketku  &=&  \sum_{\ell = 1}^n  k_{\llb_j} \ketku, \nl
p_j \ketku  &=& M \gamma u_j \ketku, \\
H \ketku &=& E(\{\kk\}; \uu) \ketku,
\eens
where $\gamma = \gamma(\uu)$ is given by (\ref{gamma}). The energy 
of the $n$-quanta state is
\be
E(\{\kk\}; \uu)= \sum_{\ell = 1}^n  \w_{\kk_\llb} +
\sqrt{ M^2 + (M\gamma\uu -\sum_{\ell = 1}^n \kk_\llb)^2}.
\label{e_n}
\ee
In particular, the energy of the ground state is
\be
E(\uu) = \sqrt{M^2 + (M\gamma\uu)^2} = M\gamma,
\label{eps0}
\ee
which is recognized as the energy of a particle of mass $M$ and 
velocity $\uu$, which is how we have modeled the observer. The energy 
of a one-quantum state is
\be
E(\kk; \uu) = \w_\kk + \sqrt{M^2 + (M\gamma\uu - \kk)^2}.
\label{epsk}
\ee
If we set $\uu = \zero$, the energy of the one-quantum state reduces to
\be
E(\kk; \zero) = \w_\kk + \sqrt{M^2 + \kk^2}.
\label{e_0}
\ee
Thus it appears as the observer's mass is $\kk$-dependent, and equal to
$M(\kk) = \sqrt{M^2 + \kk^2}$. 
On the other hand, in a situation where the observer's momentum equals
the momentum of the quantum, i.e. $M\gamma\uu = \kk$,
the energy reaches its minimum value
\be
E(\kk; \uu) = \w_\kk + M.
\ee
These formulas also apply to the multi-quanta state (\ref{e_n}), 
provided that we interpret $\kk$ as the total momentum of all
quanta, i.e. $\kk = \sum_{\ell=1}^n \kk_\llb$.

\subsection{ Non-relativistic limit }

Let us now specialize to the case that observer is much heavier than
the energy of the quanta, i.e. the limit $M \to \infty$. The 
single-quantum energy (\ref{epsk}) becomes
\be
E(\kk; \uu) \approx M\gamma + \w_\kk - \uu\cdot\kk
+ {1\/2M\gamma}(\kk^2 - 2(\uu\cdot\kk)^2) 
+ O \bigg({|\kk|^3\/M^2} \bigg).
\ee
The first term is simply the relativistic energy (\ref{eps0}) of the 
point-like observer. The next two terms are proportional to $\hbar$ and
independent of the observer's mass. They express that an observer
that moves with velocity $\uu$ experiences a Doppler shift. Relative to
the moving observer, the frequency of a quantum with wave vector $\kk$
is shifted to
\be
\w_\kk \longrightarrow \w_\kk - \uu\cdot\kk.
\ee
The final term is a genuinely new effect which is due to the observer 
dependence of relative fields. It asserts that in addition to the Doppler 
shift, the frequency of the one-quantum state acquires an extra shift
\be
{1\/2M\gamma}(\kk^2 - 2(\uu\cdot\kk)^2).
\ee
Note that this effect is present even if the observer is not moving 
relative to the global origin. When $\uu = 0$, the energy of a 
single-quantum state is $E(\kk; \zero) = M + \eps_\kk$,
where
\be
\eps_\kk \approx \w_\kk + {|\kk|^2\/2M}.
\label{e_k}
\ee
The energy of the quanta is not additive. The total energy
of the observer and quanta with wave-vectors $\kk_1$ and $\kk_2$ is 
\be
E(\kk_1, \kk_2; \zero) \approx M + \eps_{\kk_1} + \eps_{\kk_2}
+ {1\/M} \kk_1\cdot\kk_2.
\label{inter}
\ee
The last interference term appears as an interaction between the two 
quanta, and is a genuine QJT effect; it disappears in the QFT limit 
$M \to \infty$.

\subsection{ Other observer states }

So far we assumed that the observer is in a velocity eigenstate $\ket\uu$.
Since the observer obeys the rules of quantum mechanics, this means that
its position is entirely unknown. A general observer state $\ket f$ is
some linear superposition of velocity eigenstates, {\em viz.}
\be
\ket f = \int d^d\!u\, f(\uu) \ket\uu,
\ee
and the Fock space can be constructed by applying creation operators to
the vacuum $\ket f$.
In particular, a position eigenstate is the linear superposition
\be
\ket\xx = \int d^d\!p(\uu)\, \e^{i\pp(\uu) \cdot \xx} \ket\uu,
\ee
where $\pp(\uu) = M\gamma(\uu)\uu$. When acting on a state $\ketkx$
with $n$ quanta with wave-vectors $\kk_\llb$, the Hamiltonian takes
the form $H = H_\phi + H_q$, where
\be
H_\phi &=& \sum_{\ell = 1}^n \w_{\kk_\llb}, \nle
H_q &=& \sqrt{ M^2 + (i\nabla +\sum_{\ell = 1}^n \kk_\llb)^2}.
\eens
We only consider velocity eigenstates in this article.

\section{ The free electromagnetic field }

\subsection{ Action and Hamiltonian }

We next turn to describe electromagnetism in terms of relative fields.
Here we encounter two new phenomena: gauge symmetry and 
interaction between the observer and the fields. To lighten the notation,
the subscript $R$ is suppressed, keeping in mind that all fields are
relative. 
The action consists of three terms, which describe the electromagnetic
field itself, the observer's trajectory, and the interaction between the
field and the observer. We assume that the observer is a charged
particle with charge $e$. The presence of an explicit field-observer
interaction is the main novelty in this section.

The action reads $S = S_A + S_q + S_{qA}$, where
\bes
S_A &=& - {1\/4} \iint dt\,\dxx\, F^\mn(t,\xx)F_\mn(t,\xx)\nl
&=&\iint dt\,\dxx\, \big( {1\/2} F_{0i}(t,\xx)F_{0i}(t,\xx) 
- {1\/4} F_\ij(t,\xx)F_\ij(t,\xx) \big), \nl
S_q &=& -M \int dt\, \sqrt{1 - \dot\qq^2(t)}, \\
S_{qA} &=& e \int dt\, \dot q^\mu(t) A_\mu(t,\zero) 
= e \int dt\, (A_0(t,\zero) - \dot q_i(t) A_i(t,\zero)).
\eens
As usual, the field strength is
\be
F_\mn = \d_\mu A_\nu - \d_\nu A_\mu.
\ee
In particular, 
\bes
F_{0\mu}(t,\xx) &=& \d_0 A_\mu(t,\xx) - \d_\mu A_0(t,\xx) \nle 
&=& \dot A_\mu(t,\xx) + \dot q_j \d_j A_\mu(t,\xx) - \d_\mu A_0(t,\xx).
\eens
The canonical momenta are
\bes
E_\mu(\xx) &\equiv& {\d\L\/\d\dot A_\mu(\xx)} = F_{0\mu}(\xx), \nle
p_i &\equiv& {\d L\/\d\dot q_i}
= M\gamma(\dot\qq)\dot q_i + P_i - eA_i(\zero),
\eens
where
\bes
P_i &=& \int \dxx\, \d_i A_j(\xx)F_{0j}(\xx) 
\nlb{PiE}
&=& \int \dxx\, \d_i A_j(\xx)E_j(\xx) 
= -\int \dxx\, A_j(\xx)\d_i E_j(\xx).
\eens
We postulate the nonzero canonical commutators
\bes
[A_\mu(\xx), E_\nu(\xx')] &=& -i\eta_\mn \dlt(\xx-\xx'), \nle
[q_i, p_j] &=& i\dlt_\ij.
\eens
Because of the primary constraint
\be
E_0(\xx) \approx 0,
\label{primary}
\ee
the Hamiltonian $H = H_A + H_q$ depends on an arbitrary Lagrange 
multiplier $u_1(\xx)$:
\bes
H &=& \int \dxx\, \dot A_\mu(\xx) E^\mu(\xx) + \dot q_i p_i - L
+ \int \dxx\, u_1(\xx) E_0(\xx), \nl
H_A &=& \int \dxx\, \big( {1\/2} E_iE_i + {1\/4}F_\ij F_\ij + 
E_i\d_iA_0 + u_1E_0 \big),	\\
H_q &=& \MpPA,
\eens
Hamilton's equations read
\bes
\d_0 A_i(\xx) &=& E_i(\xx) + \d_i A_0(\xx), \nl
\d_0 E_i(\xx) &=& \d_jF_{ji}(\xx) - e\dot q_i\dlt(\xx), \nl
\dot A_0(\xx) &=& -u_1(\xx), \nle
\dot E_0(\xx) &=& -\d_iE_i(\xx) + e\dlt(\xx), \nl
\dot q_i &=& {p_i - P_i + eA_i(\zero) \/\MpPA}, \nl
\dot p_i &=& 0,
\eens
where the action of $\d_0$ on relative fields is still defined by 
(\ref{d0fR2}). Time evolution of the constraint (\ref{primary}) 
gives rise to the secondary constraint (Gauss' law)
\be
J(\xx) \equiv \d_i E_i(\xx) - e\dlt(\xx) \approx 0.
\label{Gauss}
\ee
Gauss' law commutes with the Hamiltonian, since
$\d_0\dlt(\xx) = \dot q_i\d_i\dlt(\xx)$, and hence there are no further
constraints. Gauss' law allows us to add the term
$H_2 = \int \dxx\, u_2(\xx) J(\xx)$ to the Hamiltonian. This modifies
the time evolution for $A_i$ into
\be
\d_0 A_i(\xx) = E_i(\xx) + \d_i A_0(\xx) - \d_i u_2(\xx).
\ee
There are several elegant methods to deal with constrained Hamiltonian
systems. Since our interest is to exhibit the effects of observer
dependence, we simply solve the constraint by imposing the gauge 
fixing conditions
\be
A_0(\xx) = \d_i A_i(\xx) = 0,
\ee
and replacing Poisson brackets by Dirac brackets. The physical 
degrees of freedom transverse fields $A^T_i(\xx)$ and $E^T_i(\xx)$.
The Hamiltonian consists of two terms, $H = H_A + H_q$, where the
field part takes the familiar form in four dimensions:
\be
H_A = {1\/2} \int \dxx\, 
\big(  E^T_i(\xx) E^T_i(\xx) + B^T_i(\xx) B^T_i(\xx) \big),
\ee
where the magnetic field is 
$B_i = \half\eps_{ijk} F_{jk} = \eps_{ijk}\d_j A_k$.

\subsection{Fourier space}

The fundamental brackets in Fourier space,
\be
[A_i(\kk), E_j(\kk')] = i\dlt_\ij \dlt(\kk+\kk'),
\ee
imply that
\be
[E_i(\kk), B_j(\kk')] = - \eps_{ij\ell} k_\ell\dlt(\kk+\kk'),
\ee
where $B_i(\kk) = i\eps_{ij\ell} k_j A_\ell(\kk)$ are the Fourier 
components of the magnetic field. It is useful to consider the dual
magnetic field 
\be
\tilde B_i(\kk) = i\eps_{ij\ell}{k_j\/|\kk|}B_\ell(\kk)
= |\kk| \Delta_\ij(\kk) A_j(\kk),
\ee
where
\be
\Delta_\ij(\kk) =\Delta_{ji}(\kk) = \dlt_\ij - {k_i k_j\/|\kk|^2}
\ee
satisfies $k_j \Delta_\ij(\kk) = 0$. 
Introduce the oscillators
\bes
a_i(\kk) &=& {1\/\stwk}( E_i(\kk) - i\tilde B_i(k) ), \nle
a^\dagger_i(\kk) &=& {1\/\stwk}( E_i(\kk) + i\tilde B_i(k) ),
\eens
which satisfy the CCR
\be
[a_i(\kk), a^\dagger_j(\kk')] = \Delta_\ij(\kk) \dlt(\kk+\kk').
\label{aiaj}
\ee
The Gauss law constraint (\ref{Gauss}) takes the form
\bes
J(\kk) &=& k_i E_i(\kk) - e 
\label{Gaussk}\\
&=& i\sqrt{|\kk|\/2} \Big( k_i a_i(\kk) + k_i a^\dagger_i(\kk) \Big)
- e \approx 0.
\eens
It is compatible with the brackets (\ref{aiaj}), because the observer's
charge $e$ commutes with $a_i(\kk)$ and $a^\dagger_i(\kk)$.
Of the $d$ pairs of oscillators, only $d-1$ are independent. We therefore
introduce the standard polarization vectors $\eps_{i\al}(\kk)$, where
$\al,\bt$ run over the $d-1$ transverse directions. The following relations
hold:
\be
\eps_{i\al}(\kk)\eps_{i\bt}(-\kk) = \dlt_\ab, \qquad
k_i\eps_{i\al}(\kk) = 0.
\label{spin1}
\ee
The transverse oscillators,
\be
a_\al(\kk) = \eps_{i\al}(-\kk)a_i(\kk), \qquad
a^\dagger_\al(\kk) = \eps_{i\al}(-\kk)a^\dagger_i(\kk),
\ee
satisfy	a non-degenerate Heisenberg algebra,
\be
[a_\al(\kk), a^\dagger_\bt(\kk')] = \dlt_\ab \dlt(\kk+\kk').
\ee
The field part of the Hamiltonian $H_A$ can now be written as a sum of
$d-1$ independent harmonic oscillators, 
\be
H_A = \sum_{\al=1}^{d-1} 
  \int \dkk\, |\kk| a^\dagger_\al(\kk) a_\al(-\kk).
\label{HA}
\ee
Note that this property is not destroyed by the presence of the
observer's charge in (\ref{Gaussk}). The observer part is more 
interesting, since it contains the interaction term
\be
H_q = \MpPA.
\ee
To lowest order in the charge, we can write the Hamiltonian as
\be
H_q = H_q^0 + eH_q^1 + O(e^2),
\label{EMexp}
\ee
where
\bes
H_q^0 &=& \MpP, 
\label{Hq01}\\
H_q^1 &=&  {A_i(\zero)(p_i-P_i)\/\MpP}.
\eens
The $H_q^1$ term suffers from an ordering ambiguity, because $A_i(\zero)$
and $P_i$ do not commute. Since
\be
A_\al(\xx) &=& \int \dkk\, \e^{i\kk\cdot\xx} \eps_{i\al}(\kk) A_i(\kk) \nle
&=& i \int {\dkk\/\stwk} \e^{i\kk\cdot\xx} \eps_{i\al}(\kk)
 (a_\al(\kk) - a^\dagger_\al(\kk)),
\eens
we define $H_q^1$ as the normal-ordered expression
\be
H_q^1 &=& i\int {\dkk\/\stwk} \eps_{i\al}(\kk) 
 ({p_i-P_i\/H_q^0} a_\al(\kk) - a^\dagger_\al(\kk){p_i-P_i\/H_q^0}).
\label{Hq1}
\ee

\subsection{ Quantization }

Let $\ket\uu$ be the vacuum state in the presence of an observer moving 
with velocity $\uu$, and let 
\be
\ket{(\kk_\one,\al_\one),...(\kk_\nnn,\al_\nnn); \uu}
= a^\dagger_{\al_\one}(\kk_\one) ...
a^\dagger_{\al_\nnn}(\kk_\nnn) \ket\uu
\ee
denote an $n$-photon state built over it. The photons are characterized by
their momentum $\kk_\llb$ and polarization $\al_\llb$. In particular,
we will consider the one-photon state $\ket{(\kk,\al);\uu}$ and the 
two-photon state $\ket{(\kk,\al), (\kk',\bt); \uu}$.

Introduce the dual states
\be
\bra{(\kk_\one,\al_\one),...(\kk_\nnn,\al_\nnn); \uu}
= \bra\uu a_{\al_\nnn}(-\kk_\nnn) ... a_{\al_\one}(-\kk_\one).
\ee
The inner product defined by $\bracket{\uu}{\uu'} = \delta(\uu-\uu')$ 
is in general not normalized. However, one-photon states are normalized,
\be
\bracket{(\kk,\al);\uu}{(\kk',\bt);\uu'}
= \dlt(\kk-\kk')\dlt_\ab\dlt(\uu-\uu'),
\ee
as are multi-photon states where are photons are different.

The vacuum satisfies
\be
p_i\ket\uu = M\gamma(\uu) u_i\ket\uu, \qquad P_i\ket\uu = 0.
\ee
In the absense of photons, the field part of the Hamiltonian vanishes,
$H_A\ket\uu = 0$, whereas 
\bes
H_q^0\ket\uu &=& M \gamma(\uu) \ket\uu, \\
H_q^1\ket\uu &=& -i \sum_{\al=1}^{d-1} \int {\dkk\/\stwk} 
 \eps_{i\al}(\kk) u_i \ket{(\kk,\al);\uu}.
\eens
There are two nonzero matrix elements of the Hamiltonian, with one state
being the vacuum $\ket\uu$:
\bes
\bra\uu H\ket{\uu'} &=& M \gamma(\uu)\delta(\uu-\uu'), 
\nlb{EMmat1}
\bra{(\kk,\al);\uu} H \ket{\uu'} 
&=& -{ie\/\sqrt{2|\kk|}}  u_i \eps_{i\al}(\kk) \delta(\uu-\uu').
\eens
The first equation asserts that the expectation value of the energy is
$M\gamma(\uu)$, i.e. the energy of a point particle moving at speed $\uu$. 
The second element is the amplitude for creating a one-photon state from
the vacuum. It is nonzero provided that $u_i \eps_{i\al}(\kk) \neq 0$,
i.e. the photon's momentum must not be parallel to the observer's 
trajectory.

When the observer does not move, $u_i = 0$. The total energy of the 
vacuum $\ket\zero$ thus equals the observer's mass $M$, as expected. 
The second matrix element in (\ref{EMmat1}) vanishes. To have a nonzero
amplitude for photon creation when $\uu=\zero$, we need to consider
transition between one- and two-photon states. When $\uu = \zero$, 
\bes
H_A \ket{(\kk,\al);\zero} &=& |\kk| \ket{(\kk,\al);\zero}, \nl
H_q^0\ket{(\kk,\al);\zero} &=& 
\sqrt{M^2 + |\kk|^2}\ket{(\kk,\al);\zero}, \\
H_q^1\ket{(\kk,\al);\zero} &=& 
i \sum_{\bt=1}^{d-1} \int {\dkk'\/\sqrt{2|\kk'|}} 
 \eps_{i\bt}(\kk') {k_i\/\sqrt{M^2+|\kk|^2}}
 \ket{(\kk,\al),(\kk',\bt);\zero}.
\eens
Hence
\be
\bra{(\kk',\bt);\uu} H \ket{(\kk,\al);\zero} 
= (M+\eps_{\kk}) \delta(\kk-\kk') \dlt_\ab\delta(\uu),
\label{HEM}
\ee
where
\be
\eps_{\kk} = |\kk| + \sqrt{M^2 + |\kk|^2} - M
\ \approx\ |\kk| + {|\kk|^2\/2M}
\ee
is the QJT energy (\ref{e_k}) of a single free photon. 
The transition amplitude from one to two photons is
\bes
&&\bra{(\kk_\one,\bt_\one),(\kk_\two,\bt_\two);\uu} 
H \ket{(\kk,\al);\zero} \ =
\label{EM12}\\
&& =\ {ie \/\sqrt{M^2+\kk^2}} \bigg(
{1\/\sqrt{2|\kk_\one|}}\ k_i\eps_{i\bt_\one}(\kk_\one) 
\dlt(\kk - \kk_\two) \dlt_{\al\bt_\two} + \onetwo \bigg) \dlt(\uu).
\eens
Even in a frame where the observer's velocity is zero, there is a nonzero
matrix element for creating a two-photon state from a one-photon state, 
because the electromagnetic field interacts with the observer. Note
that the first term vanishes when $\kk$ is parallel to $\kk_\one$.
We will discuss the form of this amplitude further in 
section \ref{sec:rescale}, and contrast it to the corresponding object
in gravity.

\section{ Gravity }

\subsection{ Action }
Finally, we turn to gravity in four dimensions, described by the Einstein 
action 
\bes
S_G &=& {1\/16\pi G}\int\dfx\, \sqrt{\det g}\, R(g) \nle
&=& {1\/2\la^2}	\int\dfx\, \sqrt{\det g}\, R(g),
\label{Einstein}
\eens
where $G$ is Newton's constant and 
\be
\la = \sqrt{8\pi G\hbar\/c} = \sqrt{8\pi}\ell_{Pl}
\ee
is a parameter of the order of the Planck length $\ell_{Pl}$.
In QJT, we must also describe the observer. The total action is thus
$S = S_G + S_q$, where the observer part $S_q$ is the proper length 
in the presence of a non-flat metric $g_\mu(x)$:
\be
S_q = -M \int dt\, \sqrt{g_\mn(0) \dot q^\mu(t) \dot q^\nu(t)}.
\ee
Because we work with relative fields, it is $g_\mn(0)$ rather than
$g_\mn(q(t))$ which appears in this formula. 

We only consider linearized gravity, and thus assume that the metric 
can be written as $g_\mn = \eta_\mn + h_\mn$, where $h_\mn$ is small
compared to the Minkowski metric. Define the graviton field $\phi_\mn$
by
\be
\la \phi_\mn = \bar h_\mn = h_\mn - \half \eta_\mn h,
\ee
where indices are now raised and lowered by means of the Minkowski 
metric, and $h = \eta^\mn h_\mn$. The action (\ref{Einstein}) has a 
gauge symmetry of diffeomorphisms which allows us to eliminate eight 
of the ten components of $\phi_\mn = \phi_{\nu\mu}$. 
We impose the following gauge conditions:
\bes
\phi_{\mu0} &=& 0, \nl
\d_j \phi_\ij &=& 0, \qquad \hbox{(Lorentz gauge)} 
\label{gravgauge} \\
\phi_{ii} &=& 0.  \qquad \hbox{(no spin 0)} 
\eens
The Einstein action can be separated into a free and an interaction part, 
$S_G = S^0_G + \la S^1_G$. In this paper we will only be concerned 
with the free graviton action, which reads, with the gauge choices above,
\bes
S^0_G &=& \half \int \dfx\, \d_\rho \phi_\mn \d^\rho \phi^\mn \nle
&=& \half \int \dfx\, \big( \d_0 \phi_\ij \d_0 \phi_\ij
- \d_k \phi_\ij \d_k \phi_\ij \big).
\eens
In the same gauge, the observer action becomes
\bes
S_q &=& -M \int dt\, \sqrt{1 - g_\ij(\zero) \dot q_i(t) \dot q_j(t)} \nle
&=& -M \int dt\, \sqrt{1 - \dot \qq^2 
- \la \phi_\ij(\zero) \dot q_i(t) \dot q_j(t) }.
\eens

\subsection{Hamiltonian }

The canonical momenta are
\bes
\pi_\ij(\xx) &=& \d_0 \phi_\ij(\xx), \nle
p_i(t) &=& {M \/\dot\tau(t)} g_\ij(\zero) \dot q_j + P_i,
\eens
where $\d_0 = \d/\d t + \dot q_i \d/\d x_i$ as in (\ref{d0fR2}), and
\bes
P_i &=& \int \dEx\, \d_i \phi_{jk}(\xx) \d_0\phi_{jk}(\xx),
\nlb{Ptau}
\dot\tau(t) &=& \sqrt{1 - g_\ij(\zero) \dot q_i(t) \dot q_j(t)}
\eens
is the derivative of the proper time $\tau(t)$.
The Hamiltonian is of the form $H = H^0_G + \la H^1_G + H_q$, where
\bes
H^0_G &=& \half \int \dEx\, \big( \pi_\ij(\xx) \pi_\ij(\xx) 
+ \d_k \phi_\ij(\xx) \d_k \phi_\ij(\xx) \big), \nl
H_q &=&	{M\/\dot\tau} 
= \sqrt{ M^2 + (\pp-\PP)^2 + \la\phi_\ij(\zero)(p_i-P_i)(p_j-P_j) }, \nl
P_i &=& \int \dEx\, \d_i \phi_{jk}(\xx) \pi_{jk}(\xx), \\
\dot\tau &=& {M\/\sqrt{ M^2 + (\pp-\PP)^2 
+ \la\phi_\ij(\zero)(p_i-P_i)(p_j-P_j)}},
\eens
and the interaction Hamiltonian $H^1_G$ describes graviton-graviton
scattering. These definitions of $P_i$ and $\dot\tau$ are compatible with 
(\ref{Ptau}).

\subsection{ Oscillators }

The gauge choice (\ref{gravgauge}) means that the Fourier modes 
$\phi_\ij(\kk)$ and $\pi_\ij(\kk)$ are not independent; they are 
subject to the conditions
\be
k_j \phi_\ij(\kk) = \phi_{ii}(\kk) =
k_j \pi_\ij(\kk) = \pi_{ii}(\kk) = 0.
\label{kgauge}
\ee
The CCR read
\be
[\phi_\ij(\kk), \pi_{\ell m}(\kk')] 
= {i\/2} ( \dlt_{i\ell} \dlt_{jm} + \dlt_{j\ell} \dlt_{im} )
\dlt(\kk+\kk'), 
\ee
up to terms needed ensure compatibility with the gauge conditions.

In analogy with the electromagnetic case (\ref{spin1}), we introduce 
spin-2 polarization tensors $\eps_{ij\al}(\kk)$, $\al = 1,2$, which
satisfy
\bes
\eps_{ij\al}(\kk) \eps_{ij\bt}(-\kk) = \dlt_\ab, &\qquad&
\eps_{ij\al}(\kk) = \eps_{ji\al}(\kk), \nle
k_j \eps_{ij\al}(\kk) = 0, &&
\eps_{ii\al}(\kk) = 0.
\eens
The independent spin-2 oscillators, labelled by $\al = 1,2$, 
\bes
a_\al(\kk) &=& {1\/\stwk}\eps_{ij\al}(-\kk)
 ( \pi_\ij(\kk) - i|\kk|\phi_\ij(\kk) ), \nle
a^\dagger_\al(\kk) &=& {1\/\stwk}\eps_{ij\al}(-\kk)
 ( \pi_\ij(\kk) + i|\kk|\phi_\ij(\kk) ),
\eens
are subject to the nonzero CCR
\be
[a_\al(\kk), a^\dagger_\bt(\kk')] = \dlt_\ab \dlt(\kk+\kk').
\ee

The free field part of the Hamiltonian becomes a sum over non-interacting
gravitons,
\be
H^0_G = \int \dEx\, |\kk| a^\dagger_\al(\kk) a_\al(-\kk),
\ee
whereas the observer part can be expanded in a power series in $\la$,
$H_q = H^0_q + \la H^1_q + O(\la^2)$, where
\bes
H^0_q &=& \sqrt{M^2 + (\pp-\PP)^2} 
\equiv \sqrt{M^2 + (p_i-P_i)(p_i-P_i)}, \nl
H^1_q &=& {1\/2H^0_q} \phi_\ij(\zero) (p_i-P_i)(p_j-P_j).
\ees
Again we have an ordering ambiguity because the graviton oscillators
do not commute with
\be
P_i = \int \dEk\, k_i a^\dagger_\al(\kk) a_\al(-\kk).
\ee
We choose to normal order the interaction Hamiltonian, i.e.
\bes
H_q^1 &=& i \sum_{\al=1}^2\int {\dEk\/\stwk} \eps_{ij\al}(\kk) 
 \bigg( {(p_i-P_i)(p_j-P_j)\/2H_q^0} a_\al(\kk) - \nl
&&\qquad -\ a^\dagger_\al(\kk){(p_i-P_i)(p_j-P_j)\/2H_q^0} \bigg).
\ees
This Hamiltonian is very similar to the electromagnetic Hamiltonian in
(\ref{Hq01}) and (\ref{Hq1}), except that the spin-2 gravitons interact 
with the tensor $(p_i-P_i)(p_j-P_j)$ rather than the vector $p_i-P_i$.

\subsection{ Quantization}

We quantize the theory in complete analogy with the electromagnetic
field in the previous section. The Hamiltonian has the following
nonzero matrix elements with the vacuum $\ket\uu$, where the
observer moves with velocity $\uu$:
\bes
\bra\uu H\ket{\uu'} &=& M \gamma(\uu)\delta(\uu-\uu'), 
\nle
\bra{(\kk,\al);\uu} H \ket{\uu'} 
&=& -{i\la\/2M\sqrt{2|\kk|}} u_iu_j \eps_{ij\al}(\kk) \delta(\uu-\uu').
\eens
The first element is the energy of the moving observer itself, and the
second element vanishes if $\uu$ is parallel to $\kk$, since
$k_j \eps_{ij\al}(\kk) = 0$. In particular, the amplitude vanishes for
a still-standing observer. 

The expectation value of the energy in a one-graviton state,
\break$\bra{(\kk',\bt);\uu} H \ket{(\kk,\al);\zero}$, 
has the same value
(\ref{HEM}) as in a one-photon state. To compute the transition amplitude
between one and two photons, we need
\bes
&&H_q^1\ket{(\kk,\al);\zero} = \\ 
&&=\ {i\/2\sqrt{M^2+\kk^2}}\sum_{\al'=1}^2\int {\dkk'\/\sqrt{2|\kk'|}} 
k_ik_j \eps_{ij\al'}(\kk') \ket{(\kk,\al),(\kk',\al');\zero}.
\eens
Hence the transition amplitude is
\bes
&&\bra{(\kk_\one,\bt_\one),(\kk_\two,\bt_\two);\uu} 
H \ket{(\kk,\al);\zero} \ =	\nl
&& =\ {i\la \/2\sqrt{M^2+\kk^2}} \bigg(
{1\/\sqrt{2|\kk_\one|}}\ k_ik_j\eps_{ij\bt_\one}(\kk_\one)
\dlt(\kk - \kk_\two) \dlt_{\al\bt_\two} \nl
&&\qquad\qquad +\ \onetwo \bigg) \dlt(\uu).
\label{GR12}
\ees
The first term vanishes if $\kk$ is parallel to $\kk_\one$, since
$k_j\eps_{ij\al}(\kk) = 0$.

\section{ Rescaling and discussion }
\label{sec:rescale}

To extract how matrix elements depend on the observer's mass, it is 
useful to isolate the $M$ dependence.
Introduce the dimensionless momentum $\kka$ by setting $\kk = M\kka$,
and rescale all other quantities with appropriate powers of $M$.
If the engineering dimension $[\psi(\kk)] = D$, define the dimensionless
quantity $\psi(\kka)$ by $\psi(\kk) = M^D \psi(\kka)$. The engineering 
dimensions of the fields in $d$ dimensions are
\be
\begin{array}{ll}
[\xx] = -1 &
[\kk] = +1 \\
{[}\dxx] = -d &
[\dkk] = +d \\
{[}\phi(\xx)] = (d-1)/2 &
[\pi(\xx)] = (d+1)/2 \\
{[}\phi(\kk)] = -(d+1)/2 &
[\pi(\kk)] = (1-d)/2 \\
{[}a_\al(\kk)] = -d/2 &
[a^\dagger_\al(\kk)] = -d/2 \\
{[}\dlt(\xx)] = d &
[\dlt(\kk)] = -d \\
{[}H] = +1 &
[\uu] = 0 \\
\end{array}
\ee

The free-field Hamiltonian given by the following expression:
\bes
H_\phi &=& {M\/2} \int \dkka\, \big( \pi(\kka)\pi(-\kka) 
+ \w_\ka^2 \phi(\kka)\phi(-\kka) \big) \nle
&=& M \int \dkka\, \w_\ka a^\dagger(\kka) a(-\kka).
\eens
The energy of the free-field $n$-quanta state, when the observer's
velocity is zero, reads
\be
H &=& M\bigg( \sum_\ell \w(\kka_\llb) + 
\sqrt{1 + (\sum_\ell \kka_\llb)^2} \bigg)
\nlb{Hfree}
&\approx& M\bigg( 1 + \sum_\ell \eps_{\kka_\llb}
+ \sum_{\ell < \ell'} \kka_\llb \cdot \kka_{\ell'} \bigg),
\eens
where
\be
\eps_{\kka} = \w_{\kka} + \half \kka^2
\label{Efree}
\ee
is the energy of a single quantum. The use of relative fields thus
results in two effects. 
\begin{enumerate}
\item
The single-quantum energy (\ref{Efree}) acquires a shift $\kka^2/2$.
\item
There is an interference term in (\ref{Hfree}), which originates from 
the shift $\pp \to \pp - \PP$ in the observer's momentum.
\end{enumerate}
Both these effects vanish in the limit $\kka \to \zero$, i.e. 
$M \to \infty$, and the QJT energy reduces to the QFT energy in this 
limit. The observer's mass thus effectively becomes a cutoff, below
which QJT reduces to QFT.

In electromagnetism and gravity we also introduced an explicit
observer-field interaction in the action. Such an interaction could
of course also be introduced within the framework of absolute fields,
but the motivation in QJT is much stronger because the observer's
position is already present in the definition of $\d_0$, and as a
quantum observable it must obey some dynamics. The new term causes
the fields to interact with the observer, giving nonzero matrix elements
between states with different numbers of quanta.
In particular, when all powers of $M$ have been extracted, the matrix 
element in (\ref{EM12}) becomes
\be
\bra{(\kka_\one,\bt_\one),(\kka_\two,\bt_\two);\uu} 
H \ket{(\kka,\al);\zero} 
=e M^{d-1\/2} F \dlt(\uu),
\ee
where
\bes
F &\equiv& F((\kka_\one,\bt_\one),(\kka_\two,\bt_\two),(\kka,\al)) \\
&=&
{i \/\sqrt{1+\kka^2}} \bigg(
{1\/\sqrt{2|\kka_\one|}}\ \ka_i\eps_{i\bt_\one}(\kka_\one) 
\dlt(\kka - \kka_\two) \dlt_{\al\bt_\two} + \onetwo \bigg)
\eens
is a dimensionless number. The analogous amplitude (\ref{GR12})
in four-dimensional gravity is
\be
\bra{(\kka_\one,\bt_\one),(\kka_\two,\bt_\two);\uu} 
H \ket{(\kka,\al);\zero} 
=\la M^2 G \dlt(\uu),
\ee
where
\bes
G &\equiv& G((\kka_\one,\bt_\one),(\kka_\two,\bt_\two),(\kka,\al)) \\
&=&
{1\/2\sqrt{1+\kka^2}} \bigg(
{1\/\sqrt{2|\kka_\one|}}\ \ka_i\ka_j\eps_{ij\bt_\one}(\kka_\one)
\dlt(\kka - \kka_\two) \dlt_{\al\bt_\two} \nl
&&\qquad\qquad +\ \onetwo \bigg)
\eens
is also a dimensionless number. Let us introduce an abbreviated
notation, where $\ket n$ stands for an $n$-quanta state, and the
momenta and polarizations are implicit. The transition amplitude from
one to two quanta can then be written in $d=3$ as
\bes
\bra2 H \ket1 = eMF\dlt(\uu), &\qquad& \hbox{(EM)}, 
\nle
\bra2 H \ket1 = (\la M) MG\dlt(\uu), &\qquad& \hbox{(Gravity)}.
\eens
Since the engineering dimension of the Hamiltonian equals one, the
rescaled matrix elements are proportional to the dimensionless numbers
$e$ and $\la M$, respectively. 

We may now make some general observations about the consequences of 
relative fields.
\begin{enumerate}
\item
QFT, or at least the correct energy levels, is recovered from QJT in the 
limit $M \to \infty$. Since this assumption is incompatible with virtual 
quanta with energy $E > M$, the $M\to\infty$ limit of QJT is essentially 
QFT with cutoff scale $M$.
\item
In electromagnetism, the field-observer interaction gives nonzero 
matrix elements between states with different numbers of photons. 
Since the amplitudes are proportional to the observer's charge $e$, 
this effect vanishes in the limit $e\to0$.
\item
We expect other nongravitational interactions to behave similarly to
electromagnetism. In particular, the effect should vanish in the limit
that the observer is uncharged.
\item
The corresponding matrix element in gravity is proportional to $\la M$.
Since the Planck length $\la$ is a universal constant, graviton number
is conserved by the observer-field interaction in the limit $M \to 0$. 
However, this limit is incompatible with the assumption $M \to \infty$ 
made at point 1 above. Hence QJT does not possess a QFT limit 
specifically for gravity.
\end{enumerate}

Only linearized gravity was considered in this paper. The interaction
part of the Hamiltonian $H^1_G$ causes graviton-graviton interactions
and problems with infinities. There is no obvious reason why passing to
relative fields should improve the situation. QJT modifies the
dispersion law for energetic quanta, but the Hamiltonian (\ref{Hfree})
still grows linearly for large momenta.

However, it seems plausible that one can construct a model where 
divergent contributions from bosonic and fermionic fields cancel.
This hope emanates from the construction in \cite{Lar04,Lar06a,Lar07},
where a recipe for cancelling divergent contributions to diff anomalies
was developed, leaving only a finite cocycle in the limit of infinite
jets.

\section{ Conclusion }

Quantum Jet Theory is an UV completion of QFT; more precisely, it
is the deformation of QFT whose deformation parameter is the
observer's mass $M$. QFT is recovered from QJT in the limit 
$G = 0$, $M \to \infty$, and general relativity is recovered in the
limit $\hbar = 0$, $M \to 0$. Since these limits are mutually
incompatible, no QFT description of gravity is possible. To
construct a consistent quantum theory of gravity, we expect that 
QJT is needed, with a finite, nonzero observer mass.

It was noted in subsection \ref{ssec:frame} that observer dependence 
is not the same as frame dependence. We typically work in 
the frame of GPS satellites; spacetime points are labelled by their
GPS coordinates. In contrast, observer dependence enters ``at the 
other side of the measuring rod''. In QJT, it is the distance between
the point (the reading of a GPS device) and the physical observer that 
is the partial observable, but in QFT it is the point itself. 
This leads to a shift in the observer's momentum: $\pp \to \pp - \PP$,
where $\PP$ is the momentum of the fields. We saw in equation (\ref{Pj=0})
that $\PP = 0$ classically, at least for the free scalar field and
probably in general. This means that using relative fields only matters
on the quantum level; the classical limits of QJT and QFT are the same.

As is discussed in detail in the companion paper \cite{Lar08b}, QJT leads
to new gauge and diff anomalies not present in QFT. This unambigously
proves that QJT is substantially different from QFT, which is positive
given that QFT is incompatible with gravity. The presence of a
diff anomaly invalidates standard claims that gravity must be
holographic; it is well known from CFT that diffeomorphism symmetry on the
circle is compatible with nontrivial correlators, but only in the
presence of a Virasoro central charge.

QJT is new physics in the sense that its predictions differ from those of
QFT, but it does not add new terms to the Lagrangian. However, several
major discoveries during the past century (special relativity, quantum
mechanics, renormalization) were not primarily about new terms in the
Lagrangian, but rather about new ways thinking about observers and
observability. QJT takes one step further in this direction by upgrading
the observer to a physical actor with quantum dynamics.

\end{document}